\documentclass[a4,prd,twocolumn,showpacs,nofootinbib,floatfix,amsmath]{revtex4}

\allowdisplaybreaks

\begin{document}

\title{Grand Unified Yukawa Matrix Ansatz: The Standard Model Fermion Mass, Quark Mixing and CP Violation Parameters}

\author{Yong-Chao Zhang}
\email{zhangyongch207@mails.gucas.ac.cn}
\author{De-Hai Zhang}
\email{dhzhang@gucas.ac.cn}

\affiliation{College of Physical Sciences, Graduate University of
Chinese Academy of Sciences, Beijing 100049, P. R. China}

\begin{abstract}
We propose a new mass matrix ansatz: At the grand unified (GU)
scale, the standard model (SM) Yukawa coupling matrix elements are
integer powers of the square root of the GU gauge coupling constant
$\varepsilon \equiv \sqrt{\alpha_{\text{GU}}}$, multiplied by order
unity random complex numbers. It relates the hierarchy of the SM
fermion masses and quark mixings to the gauge coupling constants,
greatly reducing the SM parameters, and can give good fitting
results of the SM fermion mass, quark mixing and CP violation
parameters. This is a neat but very effective ansatz.
\end{abstract}


\maketitle

In the standard model (SM), the smallness of the quark and charged lepton masses and
the quark mixing angles, and the hierarchy among them has been a
problem for decades of years. Altogether with the CP violation
phase, they constitute 13 of all the 19 free parameters in the SM.
These 13 parameters may have a common origin at extremely high
energy scales and open us a narrow window to the underlying theory
beyond the SM. The latest experimental values of the SM fermion
masses \cite{PDG} can be translated to the running masses at $M_{Z}$
with the routine and method of Ref.~\cite{xing}, wich is given in
Table~\ref{massesmz}. The three quark mixing angles and the CP
violation phase can also be easily obtained from the latest
experimental values of the four Wolfenstein parameters of the CKM
matrix \cite{PDG,Wolfenstein,CKM}:
\begin{eqnarray}
\sin\theta_{12} &=& 0.2257^{+0.0009}_{-0.0010} \;, \nonumber \\
\sin\theta_{23} &=& 0.0415^{+0.0014}_{-0.0015} \;, \nonumber \\
\sin\theta_{13} &=& 0.00359^{+0.00040}_{-0.00034} \;, \nonumber \\
|\sin\delta| &=& 0.9329^{+0.0178}_{-0.0382} \;.
\end{eqnarray}
In the SM, the neutrinos are definitely massless.

\begin{table}[!b]
\caption{\label{massesmz} The SM fermion running masses at $M_{Z}$.
}
\begin{ruledtabular}
\begin{tabular}{ccc}
$m_{u}~({\text{MeV}})$ & $m_{d}~({\text{MeV}})$ & $m_{s}~({\text{MeV}})$ \\
\hline
$1.48^{+0.47}_{-0.62}$ & $2.92^{+0.62}_{-0.93}$ & $60^{+17}_{-21}$  \\
\hline\hline
$m_{c}~({\text{GeV}})$ & $m_{b}~({\text{GeV}})$ & $m_{t}~({\text{GeV}})$ \\
\hline
$0.63^{+0.06}_{-0.07}$ & $2.89^{+0.15}_{-0.08}$ & $170.1\pm2.5$ \\
\hline\hline
$m_{e}~({\text{MeV}})$ & $m_{\mu}~({\text{MeV}})$ & $m_{\tau}~({\text{GeV}})$ \\
\hline
$0.490^{0.000}_{0.000}$ & $103^{0}_{0}$ & $1.746^{0.000}_{0.000}$ \\
\end{tabular}
\end{ruledtabular}
\end{table}

Stitching proper mass matrices is an effective way to reduce the SM
parameters, explore the fundamental physics beyond the SM and give
meaningful predictions, and many different structures of mass
matrices have been proposed and researched intensively
\cite{ansatze-FN1,ansatze-FN2,ansatze,ansatze and
relations-GJ,ansatze and relations}. What's more, many relations
among the 13 parameters are always incorporated within the fermion
mass matrices \cite{ansatze and relations-GJ,ansatze and
relations,relations}, such as the GU bottom-tauon mass unification
$m_{b}=m_{\tau}$ and the formula for the Cabibbo angle
$\tan^{2}\theta_{C}\approx m_{d}/m_{s}$.

If all the Yukawa matrix elements are of order unity at extreme high
energy scales, e.g. at the GU scale $M_{\text{GU}}$, it is
impossible to generate the hierarchical SM fermion spectroscopy
\emph{naturally}, either with the SM or the supersymmetric (SUSY)
renormalization group equations (RGEs) \cite{ansatze-FN1}. Thus the
mass matrices must be constructed hierarchically at some high energy
scale.

There is an extremely \emph{natural} small parameter at hand to
construct the hierarchical mass matrices: $\alpha_{\text{GU}} \simeq
0.04 = 0.2^{2}$. It is more convenient to use $\varepsilon \equiv
\sqrt{\alpha_{\text{GU}}} \simeq 0.2$ instead of
$\alpha_{\text{GU}}$, as 0.04 is too small. We can guess the
smallness and hierarchy of the SM fermion masses and the three SM
gauge coupling constants have common origin at $M_{\text{GU}}$: They
are related by $\alpha_{\text{GU}} \equiv \varepsilon^{2}$.
Coincidentally, $\varepsilon \equiv \sqrt{\alpha_{\text{GU}}} \simeq
\lambda$, one parameter in Wolfenstein parameterization of the CKM
matrix \cite{Wolfenstein}. We yet stress that the mass matrices are
constructed by $\varepsilon \equiv \sqrt{\alpha_{\text{GU}}}$ but
not by $\lambda$, which, reflecting the hierarchy of the quark
mixings, can be deduced from proper hierarchical quark matrices.

For example,  at $M_{\text{GU}}$,
\begin{equation}
Y_{U}(M_{\text{GU}}) \sim
\begin{pmatrix}
\varepsilon^{n_{u}}&\varepsilon^{n_{uc}}&\varepsilon^{n_{ut}} \\
\varepsilon^{n_{cu}}&\varepsilon^{n_{c}}&\varepsilon^{n_{ct}} \\
\varepsilon^{n_{tu}}&\varepsilon^{n_{tc}}&\varepsilon^{n_{t}}
\end{pmatrix}\;,
\end{equation}
where $n_{u,\,c,\,t}$, $n_{uc,\,cu,\,ut,\,tu,\,ct,\,tc}$ are all
non-negative real numbers. The elements must be multiplied by phase
factors to generate CP violation, it is a more natural choice to
multiply order unity random complex numbers
\cite{ansatze-FN1,ansatze-FN2,F-theory,naturalcpv,landscape,NS}.
Specifically, we propose the arguments of the complex numbers be
distributed evenly, but the \emph{logarithms} of the magnitudes of
the complex numbers be of \emph{normal} distribution, where the
logarithmic operation guarantees that the magnitudes of the complex
numbers are positive definite. If all the exponent numbers of the
mass matrices are only non-negative real numbers, with no other
constraints, they will be viewed as input parameters. Then there
will be more input parameters than the 13 SM phenomenological
parameters to fit, which is opposite to the initial desire to reduce
the SM parameters. Alternatively, if they are all non-negative
\emph{integers}, then these integers can be viewed as factors from
fundamental theories, just as the factor $3$ in the Georgi-Jarlskog
mass relation \cite{ansatze and relations-GJ} (if we consider the
matrices are constructed by powers of $\alpha_{\text{GU}}$, then the
exponent numbers are non-negative half-integers). For example, they
can be deduced from F-theory GUTs \cite{F-theory}. It is rational to
suppose further that $n_{uc}=n_{cu}$, $n_{ut}=n_{tu}$,
$n_{ct}=n_{tc}$, etc., leaving only 6 independent non-negative
integers for each mass matrix. In this way, the fermion masses,
quark mixings and CP violation, altogether 13 parameters, can be
possibly deduced from proper matrix structures of the single
parameter $\varepsilon \equiv \sqrt{\alpha_{\text{GU}}}$, and the GU
gauge coupling constant $\alpha_{\text{GU}}$ naturally generates the
three SM gauge coupling constants. The free parameters in the SM are
greatly reduced.

Choi \cite{Choi} has shown that GUTs with contorted multiplets of
fermions are also available, e.g. pairing of quarks and leptons from
different generations such as $(u,d) \oplus (\nu_{\mu},\mu)$ and
pairing of different left- and right-handed fermions such as
$(\nu_{e},e)_{L} \oplus \mu_{R}$. General and contorted Yukawa
matrices may lead to different values of observable quantities. For
simplicity, we do not consider matrices with contorted structures.

The goodness of fit (GoF) is defined to be \cite{GoF}
\begin{equation}\label{GoF}
\chi ^{2} = \sum_{i} \left( \ln \frac{q_{i\text{ calc}}}{q_{i\text{ expt}}}
\right)^{2} \;,
\end{equation}
where $q_{i\text{ calc}}$ and $q_{i\text{ expt}}$ are respectively quantities from
theoretical calculations and experimental observations.

Using the minimal supersymmetric standard model (MSSM) RGEs, the
Yukawa matrices are evolved from $M_{\text{GU}}$ to the SUSY scale
$M_{\text{SUSY}}$. After the SUSY particles decouple, the matrices
are evolved further to lower energy scales with the SM RGEs. The
one-loop SM and MSSM $\beta$-functions needed can be found in
Refs.~\cite{RGE1,RGE2}. At $M_{Z}$, the SM fermion mass, quark
mixing and CP violation parameters are extracted to fit the
experimental values. For simplicity, we ignore the possible effects
on these SM parameters from the decoupling of the SUSY particles,
the running of the $\tan\beta = v_{u}/v_{d}$ and the transition
between different renormalization schemes, and assume the SUSY
particles all decouple at a common scale: $M_{\text{SUSY}}$
\cite{xing,AS}. We will use mainly $M_{\text{SUSY}}=1\text{ TeV}$ in
the following calculation. However, $M_{\text{SUSY}}$ has influence
on the values of $M_{\text{SUSY}}$ and $\alpha^{-1}_{\text{GU}}$ to
some extent, which is obvious in Table~\ref{gutmalpha}.
\begin{table}[!]
\caption{\label{gutmalpha}$M_{\text{GU}}$ and $\alpha_{\text{GU}}$
for different $M_{\text{SUSY}}$. }
\begin{ruledtabular}
\begin{tabular}{cccc}
$M_{\text{SUSY}}$ & $M_{t}$ & 1 TeV & 5 TeV \\
\hline
$M_{\text{GU}}$ ($10^{16} \text{ GeV}$) & 1.87  & 1.42  & 1.11 \\
$\alpha^{-1}_{\text{GU}}$      & 24.71 & 25.87 & 26.89 \\
\end{tabular}
\end{ruledtabular}
\end{table}

To run the matrices below $M_{\text{SUGY}}$, we must first determine
$\tan\beta$. If we suppose $y_{t}\sim y_{b}\sim y_{\tau} \sim 1$ at
$M_{\text{GU}}$, $\tan\beta$ must be large. To obtain $\tan\beta$
conveniently, the three Yukawa matrices can be approximated
\begin{equation}
Y_{U,\,D,\,E} (M_{\text{GU}}) \sim
\begin{pmatrix}
0&0&0 \\ 0&0&0 \\0&0&1
\end{pmatrix},
\end{equation}
with order unity random complex numbers multiplying the $(3,3)$
elements. When the matrices are evolved from $M_{\text{GU}}$ to
$M_{Z}$ and the third generation masses are compared to the
experimental values given in Table~\ref{massesmz}, we obtain
reliably $\tan\beta\approx55$, ignoring the effect from
$M_{\text{SUGY}}$. The value of $\tan\beta$ is consistent with the
solution given in Ref.~\cite{RGE1}.

We find that when all the exponent integers of the three SM Yukawa
coupling matrices take proper values at $M_{\text{GU}}$, it is
possible to obtain good fitting results of the nine SM fermion
masses. For example, when
\begin{eqnarray}\label{integers}
Y_{U}(M_{\text{GU}}) &\sim&
\begin{pmatrix}
\varepsilon^{8}&\varepsilon^{6}&\varepsilon^{4} \\
\varepsilon^{6}&\varepsilon^{4}&\varepsilon^{3} \\
\varepsilon^{4}&\varepsilon^{3}&1
\end{pmatrix}, \nonumber \\
Y_{D}(M_{\text{GU}}) &\sim&
\begin{pmatrix}
\varepsilon^{5}&\varepsilon^{4}&\varepsilon^{3} \\
\varepsilon^{4}&\varepsilon^{3}&\varepsilon^{2} \\
\varepsilon^{3}&\varepsilon^{2}&1
\end{pmatrix}, \nonumber \\
Y_{E}(M_{\text{GU}}) &\sim&
\begin{pmatrix}
\varepsilon^{5}&\varepsilon^{4}&\varepsilon^{3} \\
\varepsilon^{4}&\varepsilon^{3}&\varepsilon \\
\varepsilon^{3}&\varepsilon&1
\end{pmatrix},
\end{eqnarray}
we will get at $M_{Z}$
\begin{eqnarray}
&&m_{u}=2.14 \text{ MeV},\,m_{c}=0.599 \text{ GeV},\,m_{t}=176.9
\text{ GeV}, \nonumber \\
&&m_{d}=3.21 \text{ MeV},\,m_{s}=55.6 \text{ MeV},\,m_{b}=3.08
\text{ GeV}, \nonumber \\
&&m_{e}=0.951 \text{ MeV},\,m_{\mu}=136 \text{ MeV},\,m_{\tau}=1.84
\text{ GeV}.
\end{eqnarray}
Compared with the experimental values in Table~\ref{massesmz}, using
Eq.~(\ref{GoF}), we can obtain a small GoF $\chi^2=0.68$ for all the nine
SM fermion masses. When the experimental errors are taken into
consideration, the GoF is even smaller: $\chi^2=0.53$.

Matrices with vanishing elements at $M_{\text{GU}}$ are also
possible choice \cite{ansatze,ansatze and relations-GJ,ansatze and
relations}. The exponent integers of the vanishing elements are
infinity. For example, if we take
\begin{equation}
Y_{U}(M_{\text{GU}}) \sim
\begin{pmatrix}
\varepsilon^{8}&0&\varepsilon^{4} \\
0&\varepsilon^{4}&0 \\
\varepsilon^{4}&0&1
\end{pmatrix},
\end{equation}
the up-type quark masses at $M_{Z}$ are
\begin{equation}
m_{u}=1.55 \text{ MeV},\,m_{c}=0.597 \text{ GeV},\,m_{t}=176.9
\text{ GeV} \nonumber
\end{equation}
with a very small GoF $\chi^2=7.1\times10^{-3}$ for the three masses.
Examples of down-type quark and charged lepton matrices with
vanishing elements are as follows:
\begin{equation}
Y_{D}(M_{\text{GU}}) \sim
\begin{pmatrix}
0&\varepsilon^{4}&0 \\
\varepsilon^{4}&\varepsilon^{3}&\varepsilon^{2} \\
0&\varepsilon^{2}&1
\end{pmatrix},
\end{equation}
with mass eigenvalues at $M_{Z}$
\begin{equation}
m_{d}=2.21 \text{ MeV},\,m_{s}=55.4 \text{ MeV},\,m_{b}=3.08 \text{
GeV} \nonumber
\end{equation}
and a very small GoF $\chi^2=8.8\times10^{-3}$ for the three down-type
quark masses;
\begin{equation}
Y_{E}(M_{\text{GU}}) \sim
\begin{pmatrix}
0&\varepsilon^{4}&\varepsilon^{3} \\
\varepsilon^{4}&\varepsilon^{2}&0 \\
\varepsilon^{3}&0&1
\end{pmatrix},
\end{equation}
with mass eigenvalues at $M_{Z}$
\begin{equation}
m_{e}=0.337 \text{ MeV},\,m_{\mu}=125 \text{ MeV},\,m_{\tau}=1.77
\text{ GeV} \nonumber
\end{equation}
and a small GoF $\chi^2=0.18$ for the three charged leptons masses.
There are no constraints on the structures of the mass matrices; in
principle we can construct any textures that lead to well fitted
fermion masses. We can even construct matrices at $M_{\text{GU}}$,
whose first and second family fermion masses are both completely
from mixing with other generations. Here is an example:
\begin{equation}
Y_{U}(M_{\text{GU}}) \sim
\begin{pmatrix}
0&\varepsilon^{6}&\varepsilon^{5} \\
\varepsilon^{6}&0&\varepsilon^{2} \\
\varepsilon^{5}&\varepsilon^{2}&1
\end{pmatrix}.
\end{equation}
Its mass eigenvalues at $M_{Z}$ are
\begin{equation}
m_{u}=1.38 \text{ MeV},\, m_{c}=0.821 \text{ GeV},\, m_{t}=176.3
\text{ GeV} \nonumber
\end{equation}
with a very small GoF $\chi^2=0.073$ for the three masses.

It is worth noting that $M_{\text{SUSY}}$ has considerable effect on
the SM fermion spectroscopy. All the masses above are obtained with
$M_{\text{SUSY}}=\text{1 TeV}$. For the matrices in
Eq.~(\ref{integers}), if we take $M_{\text{SUSY}}= M_{t}$,
\begin{equation}
m_{u}=2.62 \text{ MeV},\, m_{c}=0.673 \text{ GeV},\, m_{t}=173.8
\text{ GeV};
\end{equation}
if $M_{\text{SUSY}}=\text{5 TeV}$,
\begin{equation}
m_{u}=1.80 \text{ MeV},\, m_{c}=0.543 \text{ GeV},\, m_{t}=179.3
\text{ GeV}.
\end{equation}
Compared the up-type quark mass eigenvalues with different
$M_{\text{SUSY}}$, the effect on the first and second generation
masses is obvious. This is a general phenomenology for the quark and
charged lepton masses, whatever structure the fermion mass matrices
are at $M_{\text{GU}}$.

To obtain well fitted quark mixing and CP violation parameters, we
must pare suitable up- and down-type quark matrices. Not all paring
of good fitting up- and down-type quark matrices lead to well fitted
quark mixings and CP violation, such as the first two matrices of
Eq.~(\ref{integers}) which would give much larger
$\sin\theta_{13}=0.0136$ than the experimental value
$\sin\theta_{13}=0.00359$. When we choose
\begin{eqnarray}\label{ckm1}
Y_{U}(M_{\text{GU}}) &\sim&
\begin{pmatrix}
\varepsilon^{8}&\varepsilon^{6}&\varepsilon^{4} \\
\varepsilon^{6}&\varepsilon^{4}&\varepsilon^{3} \\
\varepsilon^{4}&\varepsilon^{3}&1
\end{pmatrix}, \nonumber \\
Y_{D}(M_{\text{GU}}) &\sim&
\begin{pmatrix}
\varepsilon^{5}&\varepsilon^{4}&\varepsilon^{4} \\
\varepsilon^{4}&\varepsilon^{3}&\varepsilon^{3} \\
\varepsilon^{4}&\varepsilon^{3}&1
\end{pmatrix},
\end{eqnarray}
we get at $M_{Z}$
\begin{eqnarray} \label{ckm-eigenvalues2}
&&\sin\theta_{12}=0.227,\, \sin\theta_{23}=0.0189,\,
\sin\theta_{13}=0.00402,\nonumber \\
&&|\sin\delta|=0.651
\end{eqnarray}
with a GoF $\chi^2=0.76$ for the four parameters. Quark matrices with
vanishing elements are also available. For example,
\begin{eqnarray}\label{ckm2}
Y_{U}(M_{\text{GU}}) &\sim&
\begin{pmatrix}
0&0&\varepsilon^{4} \\
0&\varepsilon^{4}&\varepsilon^{3} \\
\varepsilon^{4}&\varepsilon^{3}&1
\end{pmatrix}, \nonumber \\
Y_{D}(M_{\text{GU}}) &\sim&
\begin{pmatrix}
0&\varepsilon^{4}&0 \\
\varepsilon^{4}&\varepsilon^{3}&\varepsilon^{3} \\
0&\varepsilon^{3}&1
\end{pmatrix}
\end{eqnarray}
lead to
\begin{eqnarray}
&&\sin\theta_{12}=0.219,\, \sin\theta_{23}=0.0186,\,
\sin\theta_{13}=0.00257,\nonumber \\
&&|\sin\delta|=0.649 \nonumber
\end{eqnarray}
at $M_{Z}$ with a GoF $\chi^2=0.89$ for the four parameters. All the
mass eigenvalues of the quark matrices in Eq.~(\ref{ckm1}) and
Eq.~(\ref{ckm2}) are well consistent with experiments.

$M_{\text{SUSY}}$ also has possible effect on the quark mixing and
CP violation parameters. For the quark matrices in Eq.~(\ref{ckm1}),
if $M_{\text{SUSY}}= M_{t}$,
\begin{eqnarray} \label{ckm-eigenvalues1}
&&\sin\theta_{12}=0.231,\, \sin\theta_{23}=0.0211,\,
\sin\theta_{13}=0.00458,\nonumber \\
&&|\sin\delta|=0.651
\end{eqnarray}
with a GoF $\chi^2=0.65$ for the four parameters; if
$M_{\text{SUSY}}=\text{5 TeV}$,
\begin{eqnarray} \label{ckm-eigenvalues3}
&&\sin\theta_{12}=0.223,\, \sin\theta_{23}=0.0173,\,
\sin\theta_{13}=0.00359,\nonumber \\
&&|\sin\delta|=0.651
\end{eqnarray}
with a GoF $\chi^2=0.90$ for the four parameters. Compared the four
parameters in Eq.~(\ref{ckm-eigenvalues1}), (\ref{ckm-eigenvalues2})
and (\ref{ckm-eigenvalues3}), the effect of $M_{\text{SUSY}}$ on
$\sin\theta_{12}$ and $\sin\theta_{23}$ is small, the effect on
$\sin\theta_{13}$ is comparatively large, but $|\sin\delta|$ is
hardly subject to $M_{\text{SUSY}}$. It is almost the same when the
quark matrices are of other structures. The hierarchical structure
of the CKM matrix, which is reflected by the powers of $\lambda$ in
the Wolfenstein parameterization, can obtain naturally from proper
quark matrices, as we stated above.

In conclusion, we have proposed a new ansatz for the SM fermion mass
and mixing generation: At $M_{\text{GU}}$ the Yukawa coupling
matrices are constructed by non-negative integer powers of
$\varepsilon \equiv \sqrt{\alpha_{\text{GU}}}$, i.e. by non-negative
half-integer powers of $\alpha_{\text{GU}}$, with matrix elements
multiplied by order unity random complex numbers. By all the
examples given above, we have demonstrated that this is a neat but
extremely effective ansatz: It not only relates the hierarchy of SM
fermion masses and quark mixings to the gauge coupling constants,
reducing the number of the SM parameters to the utmost extent, but
can give well fitted SM fermion mass, quark mixing and CP violation
parameters. It could also apply to the neutrino mass and oscillation
problem.

\end{document}